\documentclass[11pt]{article}

\usepackage{amsmath}
\usepackage{amssymb}
\usepackage{graphicx}
\usepackage{color}
\usepackage{amsthm}
\usepackage{amsfonts}
\usepackage{amscd}
\usepackage{mathrsfs}
\usepackage{bm}
\usepackage{enumerate}
\usepackage{algorithmic, algorithm}

\makeatletter

\usepackage{amsthm}\usepackage{amsfonts}\usepackage{amscd}\usepackage{mathrsfs}\usepackage{bm}\usepackage{enumerate}

\newcommand{\D}{\displaystyle}

\newcommand{\e}{\epsilon}

\newcommand{\ot}{\overline{\theta}}

\makeatother

\begin{document}

\title{Extracting a shape function for a signal with intra-wave frequency modulation}

\author{Thomas Y. Hou,\thanks{Applied and Comput. Math, MC 9-94, Caltech,
Pasadena, CA 91125. {\it Email: hou@cms.caltech.edu.}} \and
Zuoqiang Shi\thanks{Mathematical Sciences Center, Tsinghua University, Beijing, China, 100084. 
{\it Email: zqshi@math.tsinghua.edu.cn.}} }

\maketitle

\begin{abstract}
In this paper, we consider signals with intra-wave frequency modulation. 
To handle this kind of signals effectively, we generalize our data-driven time-frequency analysis by using a shape function to describe the intra-wave frequency modulation. The idea of using a shape function in time-frequency analysis was first proposed by Wu in \cite{Wu11}. A shape function 
could be any $2\pi$-periodic function. Based on this model, we propose to solve an optimization problem to extract the shape function. 
By exploring the fact that
$s$ is a periodic function of $\theta$, we can identify certain
low rank structure of the signal. This structure enables us to extract 
the shape function from the signal. To test the robustness of our method, we apply our method on several synthetic and real signals. The results 
are very encouraging.  
\end{abstract}


\section{Introduction}

Nowadays, data play a more and more important role in our life. At the same time, the scientific community face a challenging problem: 
how to efficiently extract useful information from massive amount of data. In many real world problems, especially in engineering and physical 
problems, frequencies of the signal are usually very useful to help us understand the underlying physical mechanism.
Hence, many time frequency analysis methods have been developed, for instance, the windowed Fourier transform, the wavelet transform
\cite{Daub92, Mallat09}, the Wigner-Ville distribution \cite{Flandrin99}, etc. In recent years, an adaptive time frequency analysis method, the
Empirical Mode Decomposition (EMD) method \cite{Huang98,WH09} was developed. 
This method provides an efficient adaptive method to extract frequency information. The EMD method has found to be very useful in many applications. 
EMD method is purely empirical, it still lacks a rigorous mathematical foundation. 
Recently, several methods have been proposed attempting to provide a mathematical foundation for EMD method,
see e.g. the synchrosqueezed wavelet transform \cite{DLW11}, the Empirical wavelet transform \cite{DZ13}, the variational mode decomposition \cite{Gilles13}.

In the last few years, inspired by the EMD method and compressive sensing \cite{CRT06a,Candes-Tao06,Dnh06},
we proposed a novel time-frequency analysis method based on the sparsest time-frequency representation of multiscale data \cite{HS13}. 
In this method, the signal is decomposed into several components
\begin{eqnarray}
\label{decomp-f}
  f(t)=\sum_{j=1}^M a_j(t)\cos\theta_j(t) + r(t) ,\quad t\in \mathbb{R} ,
\end{eqnarray}
where $a_j(t),\; \theta_j(t)$ are smooth functions, $\theta_j'(t)>0,\;j=1,\cdots,M$, and $r(t)$ is a small residual. We assume that
$a_j(t)$ and $\theta_j'$ are less oscillatory than $\cos\theta_j(t)$. The exact meaning of less oscillatory will be made clear later.
We call $a_j(t)\cos\theta_j(t)$ as the Intrinsic Mode Functions (IMFs) \cite{Huang98}.

One main difficulty in computing the decomposition \eqref{decomp-f} is that the decomposition is not unique.
To pick up the "best" decomposition, we proposed to decompose the signal by looking for the sparsest decomposition
by solving a nonlinear optimization problem:
 \begin{eqnarray*}
\label{opt-l0}
 \begin{array}{rcc}\vspace{-2mm}
  &\mbox{Minimize} &M\\ \vspace{2mm}
&{\scriptstyle (a_k)_{1\le k\le M}, (\theta_k)_{1\le k\le M}}&\\
&\mbox{Subject to:}&\D f=\sum_{k=1}^M a_k\cos\theta_k,\\
& &a_k\cos\theta_k\in \mathcal{D},
\end{array}
\end{eqnarray*}
where $\mathcal{D}$ is the dictionary consist of all IMFs (see \cite{HS13} for its precise definition).

To solve \eqref{opt-l0}, we proposed two algorithms. The first one is based on matching pursuit \cite{HS13} and the other
one is based basis pursuit \cite{HS14}. 
In a subsequent paper \cite{HST14}, the authors proved the convergence of their nonlinear matching pursuit algorithm for periodic data
that satisfy certain scale separation property.

Although the model \eqref{decomp-f} has been applied to a number of applications with success and the decomposition methods have been shown to be effective and efficient, there are also some applications such as the Stokes waves or some nonlinear dynamic systems for which our methods do not work well. An essential difficulty for this type of data is that the instantaneous 
frequency, $\theta'_k$, is as oscillatory as or even more oscillatory than $\cos\theta_k$, which we call intra-wave frequency modulation. 
One consequence is that the data may not have sparse representation 
in model \eqref{decomp-f}. To effectively decompose the signal with 
intra-wave frequency modulation, we need another model to retrieve the 
sparse stucture. Our approach is to introduce a periodic shape function 
\cite{Wu11} to replace 
the cosine function in \eqref{decomp-f}, then we get following model:
\begin{eqnarray}
\label{model-shape}
  f(t)=\sum_{k=1}^M a_k(t)s_k(\theta_k(t)) + r(t),
\end{eqnarray}
where $s_k$ is an unknown $2\pi$-periodic `shape function' and is 
adapted to the signal. The envelope $a_k(t)$ and the phase function 
$\theta_k(t)$ are smooth functions and are less oscillatory 
than $\cos\theta_k(t)$. We also assume that $\theta_k'(t)>0$. 

With the introduction of shape function, even the signal with intra-wave frequency modulation has a
sparse representation. But it is more difficult to find this sparse decomposition since the dictionary is 
larger than that in \eqref{decomp-f} due to the extra degree of freedom of shape functions. 

In this paper, we will introduce a method to extract the shape function for the signals with only one 
dominated shape function. First, the phase function 
is computed by our data driven time frequency analysis \cite{HS13}. Once the 
phase function is obtained, by exploring the fact that
$s$ is a periodic function of $\theta$, we can identify certain
low rank structure of the signal. This structure enables us to extract 
the shape function from the signal.

The rest of this paper is organized as follows. In Section 2, the decomposition model for data with intra-wave frequency modulation is presented. 
The details of the algorithm and the localized version are given in Section 3 and 4. We present some numerical results in Section 5. 
Some concluding remarks are made in Section 6.

\section{Models for signal with intra-wave frequency modulation}

In order to design a computational algorithm for the model \eqref{model-shape}, we first need to define the meaning of "less oscillatory". 
With a given phase function $\theta(t)$, 
we construct a linear space $V(\theta)$ which is spanned by the following basis
\begin{eqnarray}
\label{def-V}
\left\{ \left(\cos\left(\frac{k\theta}{L_\theta}\right)\right), 
\left(\sin\left(\frac{k\theta}{L_\theta}\right)\right)\right\}_{0\le k\le \lambda L_\theta},
\end{eqnarray}
where $\lambda< 1/2$ is a parameter to control the smoothness of
functions in $V(\theta,\lambda)$ and $L_\theta=(\theta(1)-\theta(0))/2\pi$ is a positive integer and $[0,1]$ is the span of 
the signal in time. And we require $a_k(t)$ and $\theta'_k(t)$ to be $V(\theta_k,\lambda)$ to enforce that they are less oscillatory than $\cos\theta_k$.

Then, the model of the signal is given as follows:
\begin{eqnarray}
\label{model:data}
  f(t)&=&\sum_{k=1}^M a_k(t)s_k(\theta_k(t)),\quad a_k,\theta'_k \in V(\theta_k)\nonumber \\
&& \text{and}~\theta_k'>0,\;
s_k~\text{is}~2\pi\text{-periodic}.
\end{eqnarray}

Corresponding to this model, we propose to solve the following optimization problem to find the sparsest decomposition 
\eqref{model:data},
\begin{eqnarray}
\label{opt:l0_shape}
  &&\min_{a_k,\theta_k,s_k}\hspace{2cm} M\\
 &&\mbox{Subject to:} \quad \sum_{k=1}^Ma_ks_k(\theta_k)=f,\; a_ks_k(\theta_k)\in \mathcal{M}.\nonumber
\end{eqnarray}
where the dictionary $\mathcal{M}$ is defined as
\begin{eqnarray}
  \label{dic-intra}
  \mathcal{M}=\left\{a_ks_k(\theta_k):\begin{array}{l} a_k, \theta'_k\in V(\theta_k), \;\theta_k'>0,\\
 s_k\; \mbox{is $2\pi$-period function} .\end{array}\right\}
\end{eqnarray}
This optimization problem is very difficult to solve. In this paper, we focus on a simpler case. We assume the signal 
is dominated by one component in $\mathcal{M}$, i.e.
\begin{eqnarray}
\label{model:1_shape}
  f(t)&=&a(t)s(\theta(t))+r(t),\quad a,\theta' \in V(\theta),\nonumber\\ 
&&\text{and}~\theta'>0,\;
s~\text{is}~2\pi\text{-periodic}.
\end{eqnarray}
Here, $r(t)$ is the residual. The residual $r(t)$ could be noise or trend or some kind of perturbation. No matter what $r(t)$ is, we assume that
it is small in amplitude compared with $a(t)s(\theta(t))$.

Using the idea of matching pursuit, the decomposition in \eqref{model:1_shape} can be obtained by solving the following optimization problem:
\begin{eqnarray}
\label{opt:mp}
&&  \min_{a,\theta,s} \;\;\|f(t)-a(t)s\left(\theta(t)\right)\|_2^2,\\
&&\mbox{subject to:} \quad a(t)s\left(\theta(t)\right)\in \mathcal{M}.\nonumber
\end{eqnarray}
 Although this optimization problem is much simpler than \eqref{opt:l0_shape}, it is still 
very difficult to solve. It is highly nonlinear. The envelope $a$, the phase function $\theta$, and the
shape function $s$ are all unknown. They are all adaptive to the data. This feature makes our method
fully adaptive to the signal, but it also introduces additional difficulty to solve the resulting optimization problem \eqref{opt:mp}.

Inspired by our previous work in the data-driven time-frequency analysis \cite{HS13}, we develop an efficient method to solve \eqref{opt:mp}. First, the phase function is computed by our data-driven time-frequency analysis \cite{HS13}. Once the 
phase function is obtained, by exploring the fact that
$s$ is a periodic function of $\theta$, we can identify certain
low rank structure of the signal. This structure enables us to extract 
the shape function from the signal. The details will be given in the next section.

\section{An efficient algorithm to compute the shape function}

First, to simplify the discussion, we assume that the phase function $\theta$ has been obtained, 
then the optimization problem \eqref{opt:mp} can be
reduced to the following problem:
\begin{eqnarray*}
&&  \min_{a,s} \|f(t)-a(t)s\left(\theta(t)\right)\|_2^2,\\
&&\mbox{subject to:} \quad a \in V(\theta),\quad s(\cdot) \;\mbox{is $2\pi$-periodic}. \nonumber
\end{eqnarray*}
Since $s$ is periodic, it can be represented by Fourier basis, 
  \begin{eqnarray}
   s(\theta)=\sum_{k=-K}^K c_ke^{ik\theta} .
  \end{eqnarray}
In the above Fourier representation, we assume that $s$ is $K$-band limited, which is a good approximation as long as 
$s$ is smooth enough and $K$ is large enough. 

Using the above representation, the optimization problem becomes
\begin{eqnarray*}
  \min_{a,c_k}\;\; \|f-a\sum_{k=-K}^K c_ke^{ik\theta}\|_{2}^2,
\mbox{subject to:}\;  a \in V(\theta). 
\end{eqnarray*}
Next, in order to further simplify above optimization problem, we replace the standard $l^2$ norm in the objective function 
to  $l^2$ norm in $\theta$ space,
\begin{eqnarray*}
  \|f\|_{2,\theta}=\left(\int f^2 d \bar{\theta}\right)^{1/2}=\left(\int f^2(t) \bar{\theta}'(t)d t\right)^{1/2}
\end{eqnarray*}
where $\ot=\theta/L_\theta$ is normalized phase function which is used as the coordinate function in the $\theta$-space and 
$L_\theta=\frac{\theta(1)-\theta(0)}{2\pi}$. In this paper, we assume that the signal lies in $[0,1]$.

Then, the above optimization problem is reduced to 
\begin{eqnarray}
\label{opt:l2_theta}
 &&\min_{a,c_k} \left\|\sum_{\omega=-\infty}^{+\infty} \widehat{f}_\theta(\omega)e^{i\omega\theta/L_\theta}-a\sum_{k=-K}^K c_ke^{ik\theta}\right\|_{2,\theta}^2,\nonumber
\\
 &&\mbox{subject to:} \quad a \in V(\theta). 
\end{eqnarray}
where $\widehat{f}_\theta$ is the Fourier coefficients of $f$ in the $\theta$-space,
\begin{eqnarray*}
  \widehat{f}_\theta(\omega)=\int_0^1 f(t)e^{-i\omega\ot(t)}\ot'(t)dt .
\end{eqnarray*}


Next, we represent the envelope $a$ by the Fourier basis in the $\theta$-space,
\begin{eqnarray}
  a=\sum_{\omega=-\infty}^{+\infty}\widehat{a}_\theta(\omega) e^{i\omega\ot}.
\end{eqnarray}
Then we have
\begin{eqnarray}
&&  a\sum_{k=-K}^K c_ke^{ik\theta}\\
&=&
\sum_{k=-K}^K c_k\left(\sum_{\omega=-\infty}^{+\infty}\widehat{a}_\theta(\omega) e^{i(\omega+kL_\theta)\ot} \right)
\nonumber\\
&=& \sum_{k=-K}^K c_k\left(\sum_{\omega=-\infty}^{+\infty}\widehat{a}_\theta(\omega-kL_\theta) 
e^{i\omega\ot} \right)\nonumber\\
&=&\sum_{\omega=-\infty}^{+\infty}\left(\sum_{k=-K}^K c_k\widehat{a}_\theta(\omega-kL_\theta) 
 \right)e^{i\omega\ot}.
\end{eqnarray}
Then, \eqref{opt:l2_theta} becomes
\begin{eqnarray*}
&&\hspace{-3mm} \min_{\widehat{a}_\theta,c_k}\left\|\sum_{\omega=-\infty}^{+\infty} \left[\widehat{f}_\theta(\omega)-
\sum_{k=-K}^K c_k\widehat{a}_\theta(\omega-kL_\theta)\right]e^{i\omega\ot}\right\|_{2,\theta}^2 \\
&&\hspace{-3mm}\mbox{subject to:} \quad a \in V(\theta). 
\end{eqnarray*} 
Then, using the well known Parsarval equality, the objective function is equal to the $l^2$ norm of the Fourier coefficients, 
which gives rise to the following equivalent optimization problem
\begin{eqnarray*}
 && \min_{\widehat{a}_\theta,c_k}\quad \sum_{\omega=-\infty}^{+\infty}
\left|\widehat{f}_\theta(\omega)-\sum_{k=-K}^K c_k\widehat{a}_\theta(\omega-kL_\theta)\right|^2,\\
&&\mbox{subject to:} \quad a \in V(\theta). 
\end{eqnarray*}
Since $a \in V(\theta)$, using the defination of $V(\theta)$, we have
 $\widehat{a}_\theta(\omega)=0,\; |\omega|\ge L_\theta/2$. Then
 \begin{eqnarray}
&&   \sum_{\omega=-\infty}^{+\infty}
\left|\widehat{f}_\theta(\omega)-\sum_{k=-K}^K c_k\widehat{a}_\theta(\omega-kL_\theta)\right|^2\nonumber\\
&=& \sum_{j=-\infty}^{+\infty} \sum_{\omega=-L_\theta/2}^{L_\theta/2-1}
\left|\widehat{f}_\theta(\omega+jL_\theta)-\sum_{k=-K}^K c_k\widehat{a}_\theta(\omega+(j-k)L_\theta)\right|^2\nonumber\\
&=&\sum_{j=-K}^{K} \sum_{\omega=-L_\theta/2}^{L_\theta/2-1}
\left|\widehat{f}_\theta(\omega+jL_\theta)- c_k\widehat{a}_\theta(\omega)\right|^2\nonumber\\
&&+\sum_{|j|>K,\atop j\in \mathbb{Z}} \sum_{\omega=-L_\theta/2}^{L_\theta/2-1}
\left|\widehat{f}_\theta(\omega+jL_\theta)\right|^2 , \nonumber
 \end{eqnarray}
where we have used the fact that if $k\ne j$, $\widehat{a}_\theta(\omega+(j-k)L_\theta)=0$ for any $-L_\theta/2
\le \omega< L_\theta/2$ in obtaining the last equality.

Using the above derivation, we have the following equivalent optimization problem,
\begin{eqnarray*}
&&  \min_{\widehat{a}_\theta,c_k}\;\; \sum_{k=-K}^K\sum_{|\omega|< L_\theta/2}| \widehat{f}_\theta(\omega+kL_\theta)-c_k\widehat{a}_\theta(\omega)|^2.
\end{eqnarray*}

Denote
$$\widehat{f}_{\theta,k}(\omega)= \left\{\begin{array}{cl}
\widehat{f}_\theta(\omega),& kL_\theta \le \omega<(k+1)L_\theta,\\
0,& \text{otherwise},
\end{array}\right.$$
and 
$$f_{\theta,k}(\ot)=\mathcal{F}^{-1}_\theta\left(\widehat{f}_{\theta,k}(\omega)\right)(\ot)=\sum_{\omega=-\infty}^{+\infty} 
\widehat{f}_{\theta,k}(\omega)e^{i\omega \ot}.$$
Then, using the Parsaval equality one more time, we need only to solve
the following equivalent problem:
\begin{eqnarray}
\label{opt:continue_svd}
  \min_{a_\theta,c_k}\;\; \sum_{k=-K}^K\| f_{\theta,k}(\ot)-c_k a_\theta(\ot)\|_2^2,
\end{eqnarray}
where $a_\theta(\ot(t))=a(t)$ is the representation of $a$ in the $\theta$-space. Fortunately, after discretization, the above optimization problem 
can be solved by singular value decomposition (SVD).

Suppose $a_\theta$ and $f_{\theta,k}$ is sampled over $\ot_j=(j-1)/N,\; j=1,\cdots ,N$ 
which is a uniform grid in the $\theta$-space. Let
\begin{eqnarray}
\label{def:matrix_f}
&&\hspace{-20mm}\mathbf{f}_{\theta,k}=\left(f_{\theta,k}(\ot_1),\cdots,f_{\theta,k}(\ot_N)\right)^t,\\ 
&&\hspace{-20mm}\mathbf{F}_\theta=\left(\text{Re}(\mathbf{f}_{\theta,0}),\cdots,\text{Re}(\mathbf{f}_{\theta,K})),
\text{Im}(\mathbf{f}_{\theta,1}),\cdots,\text{Im}(\mathbf{f}_{\theta,K})\right),\hspace{-20mm}\nonumber
\end{eqnarray}
and
\begin{eqnarray}
\label{def:vector_ac}
&& \mathbf{a}_\theta=
\left(a_\theta(\ot_1),\cdots,a_\theta(\ot_N)\right)^t,\\
&& \mathbf{c}=(\text{Re}(c_0),\cdots,\text{Re}(c_K),
\text{Im}(c_1),\cdots,\text{Im}(c_K)).\nonumber
\end{eqnarray} 
Then, the discrete version of \eqref{opt:continue_svd} is 
\begin{eqnarray}
\label{opt:svd}
\min_{\mathbf{c}\in \mathbb{R}^{2K+1},\atop \mathbf{a}_\theta\in \mathbb{R}^N}  \|\mathbf{F}_\theta-\mathbf{a}_\theta\cdot \mathbf{c}\|_{F}^2 ,
\end{eqnarray}
where $\|\cdot\|_F$ is the Forbenius norm of a matrix. It is well known that the above optimization problem can be solved by SVD. 

Suppose $$\mathbf{F}_\theta=\mathbf{U\cdot S\cdot V}$$ is the singular value decomposition of $\mathbf{F}_\theta$, 
$\text{diag}(S)=(s_1,\cdots,s_{2K+1})$ and $s_1\ge s_2\ge \cdots,\ge s_{2K+1} \ge 0$. Then the solution of the optimization problem 
\eqref{opt:svd} is 
\begin{eqnarray}
\label{eqn:sol_svd}
  \mathbf{a}_\theta=\mathbf{u}_1,\quad \mathbf{c}=\mathbf{v}_1 ,
\end{eqnarray}
where $\mathbf{u}_1$ is the first column of matrix $\mathbf{U}$ and $\mathbf{v}_1$ is the first row of matrix $\mathbf{V}$.

Summarizing above discussion, we get Algorithm \ref{alg:shape_global} to compute the shape function with given phase function. 
\begin{algorithm}
\floatname{algorithm}{Algorithm}
\caption{(Extraction of shape function)}
\label{alg:shape_global}
\begin{algorithmic}[1]
\REQUIRE Signal $\mathbf{f}=(f(t_1),\cdots,f(t_{N}))$ is sampled over $t_l,\; l=1,\cdots,N$, 
phase functions $\theta$, band width of shape function $K$.
\ENSURE  Shape function $s$ and corresponding envelope $a(t)$.
\STATE Interpolate the original signal $\mathbf{f}$ from $t_l, l=1,\cdots, N$ to a uniform grid $\ot_j=j/N,\; j=0,\cdots, N-1$ in $\theta$ space.
\begin{eqnarray}
f_{\theta}^j=\text{Interpolate}(\theta(t_l),f(t_l),\ot_j).\nonumber
\end{eqnarray}
In the computation, we use cubic spline to do interpolation.
\STATE Compute the Fourier coefficients of $f$ in $\theta$ space
\begin{eqnarray}
  \widehat{f}_\theta(\omega)=\sum_{j=0}^{N-1} f_\theta^j e^{-i\omega \ot_j},\; \omega=-\frac{N}{2},\cdots,\frac{N}{2}-1.\nonumber
\end{eqnarray}
\STATE Compute $f_{\theta,k},\; k=0,\cdots, K$,
\begin{eqnarray}
  f_{\theta,k}(\ot_j)=\sum_{\omega=(k-1/2)L_\theta}^{(k+1/2)L_\theta-1}\widehat{f}_\theta(\omega)e^{i\omega \ot_j}.\nonumber
\end{eqnarray}
\STATE Assemble the matrix $\mathbf{F}_\theta$ according to \eqref{def:matrix_f}.
\STATE Apply SVD on $\mathbf{F}_\theta$ to get the envelope $a_\theta(\ot_j), j=1,\cdots, N$ and $c_k, k=0,\cdots,K$
according to \eqref{def:vector_ac} and \eqref{eqn:sol_svd}.
\STATE Compute the shape function and the envelope function in $t$ space
\begin{eqnarray*}
  s(t_l)&=&\sum_{k=-K}^{K}c_k e^{ikt_l},\quad l=1 ,\cdots, N,\\
  a(t_l)&=&\text{Interpolate}(\ot_j,a_\theta(\ot_j),\theta(t_l)), 
\end{eqnarray*}
where $c_{-k}=c_k^*$ is the complex conjugate of $c_k$. The interpolation is also implemented by cubic 
spline.
\end{algorithmic}
\end{algorithm}

At the end of this section, we give some remarks on how to compute the phase function $\theta$. 
Notice that $s$ has the following representation  
$$s(t)=\sum_{k=-K}^Kc_ke^{ikt}=\sum_{k=1}^K\left(b_k\cos(kt)+d_k\sin(kt)\right),$$ 
where $b_k=\text{Re}(c_k),\; d_k=-\text{Im}(c_k)$. Without loss of generality, we assume that $c_0=0$. Otherwise, the constant 
part of $s$ can be absorbed into $r(t)$ in model \eqref{model:1_shape}. Then the signal $f(t)$ can be written as follows:
\begin{eqnarray}
&&  f(t)=a(t)s(\theta(t))+r(t)\nonumber\\
&=&a(t)\sum_{k=1}^{K}\left(b_k\cos(k\theta(t))+d_k\sin(k\theta(t))\right)+r(t)\nonumber\\
&=&\sum_{k=1}^K\left(a(t)\sqrt{b_k^2+d_k^2}\right)\cos(k\theta(t)+\phi_k)+r(t),\nonumber
\end{eqnarray}
where $\phi_k=\arctan\left(\frac{d_k}{b_k}\right)$.

Then the signal $f(t)$ can be seen as the signal composed by $K$ IMFs. And these IMFs satisfy the scale 
separation property. Then, we could use the method developed in \cite{HS13} to compute the phase function. 


\section{The signal with varying shape function}
\label{sec:local-shape}

In some problems, the change of the shape function is more useful than the shape function itself. To detect the change of the 
shape function, we need some local method to extract ``instantaneous'' the shape function. The simplest idea is to cut the whole signal
into small pieces and apply the method proposed in the previous section to get the shape function in each piece. Then we can get a 
series of shape functions which could show us how the shape function varies.  

Suppose we have a signal $\mathbf{f}=(f(t_1),\cdots,f(t_{N}))$ which is sampled at $t_1,\cdots,t_N$, and 
the phase functions $\theta=(\theta(t_1),\cdots,\theta(t_N))$ is also given. For each $t_m,\; m=1,\cdots,N$, we want to use the 
signal around $t_m$ to get a shape function. 

First, we extract a small piece of signals $\mathbf{f}_m$ and the phase function $\theta_m$ around $t_m$, the length of $f_m$ depends on the phase function $\theta$,
\begin{eqnarray}
    \mathbf{f}_m=\mathbf{f}_T\chi_T,\quad \theta_m=\theta_T.\nonumber
  \end{eqnarray}
where $\mathbf{f}_T=(f(t_j))_{j\in T}$ and so is $\theta_T$.
Here $T=\{1\le j\le N: |\theta(t_j)-\theta(t_m)|\le \mu \pi\} $ and
$$\chi_T=\left(\frac{1}{2}\left(1+\cos\left( \frac{t_j}{\mu}\right)\right)\right)_{j\in T}.$$
Here $\mu$ is a parameter to control the length of the segment.
In this paper, we 
choose $\mu=3$, which means that for each 
point, we localize the signal within 3 
periods to extract the shape function. 

Once we get the segments $\mathbf{f}_m$ and $\theta_m$, the shape function can be obtained by using the method in the previous section.
For each $t_m$, repeat this process, then we get a series of shape functions which could capture the change of the shape function. 

The shape functions are a series of functions. Usually, it is not easy to distinguish which part is changing. It would be very helpful if we could find an index for each shape function and this index could reflect 
the main feature of the shape function. The concept of degree of nonlinearity was proposed by Huang (lecture in the IMA Hot Topic Workshop on Trend 
and Instantaneous Frequency, September 7-9, 2011, IMA). The main idea is that the signal comes from some physical process. The main feature of the 
shape function is the nonlinearity of the underling physical process. He defined an index to measure this nonlinearity. Later, we further developed this idea by assuming that the underling process is governed by a second order ODE with polynomial nonlinearity \cite{HST14-2}. Then we formulated an optimization problem to 
calculate the coefficients and the degree of nonlinearity of the ODE. Using these techniques, we could see clearly how the shape function changes and detect the time when 
significant change occurs.

\section{Numerical results}

In this section, we will present some numrical results to demonstrate the performance of our algorithm. 

\begin{figure}
    \begin{center}
      \includegraphics[width=0.3\textwidth]{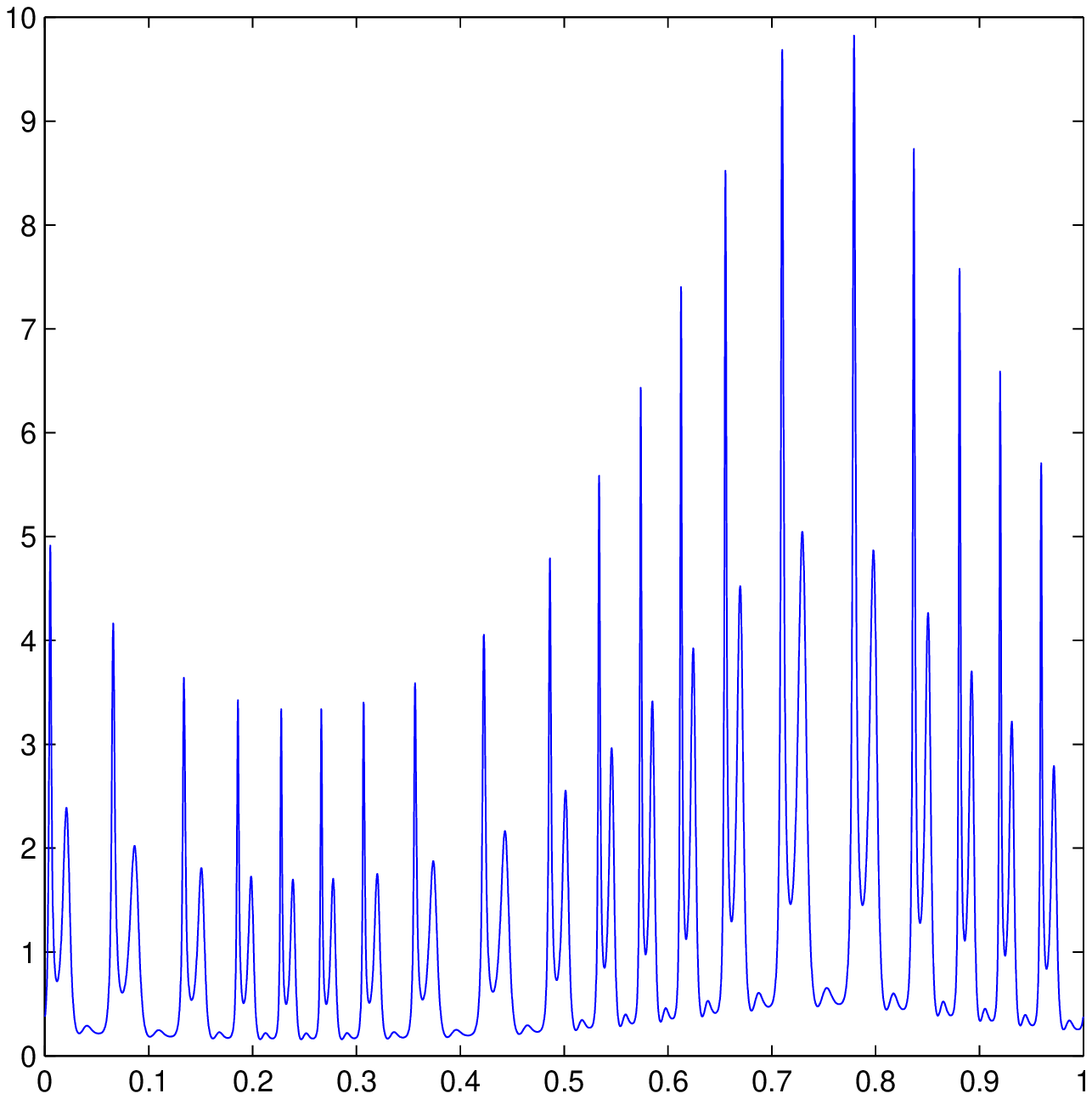}
      \includegraphics[width=0.3\textwidth]{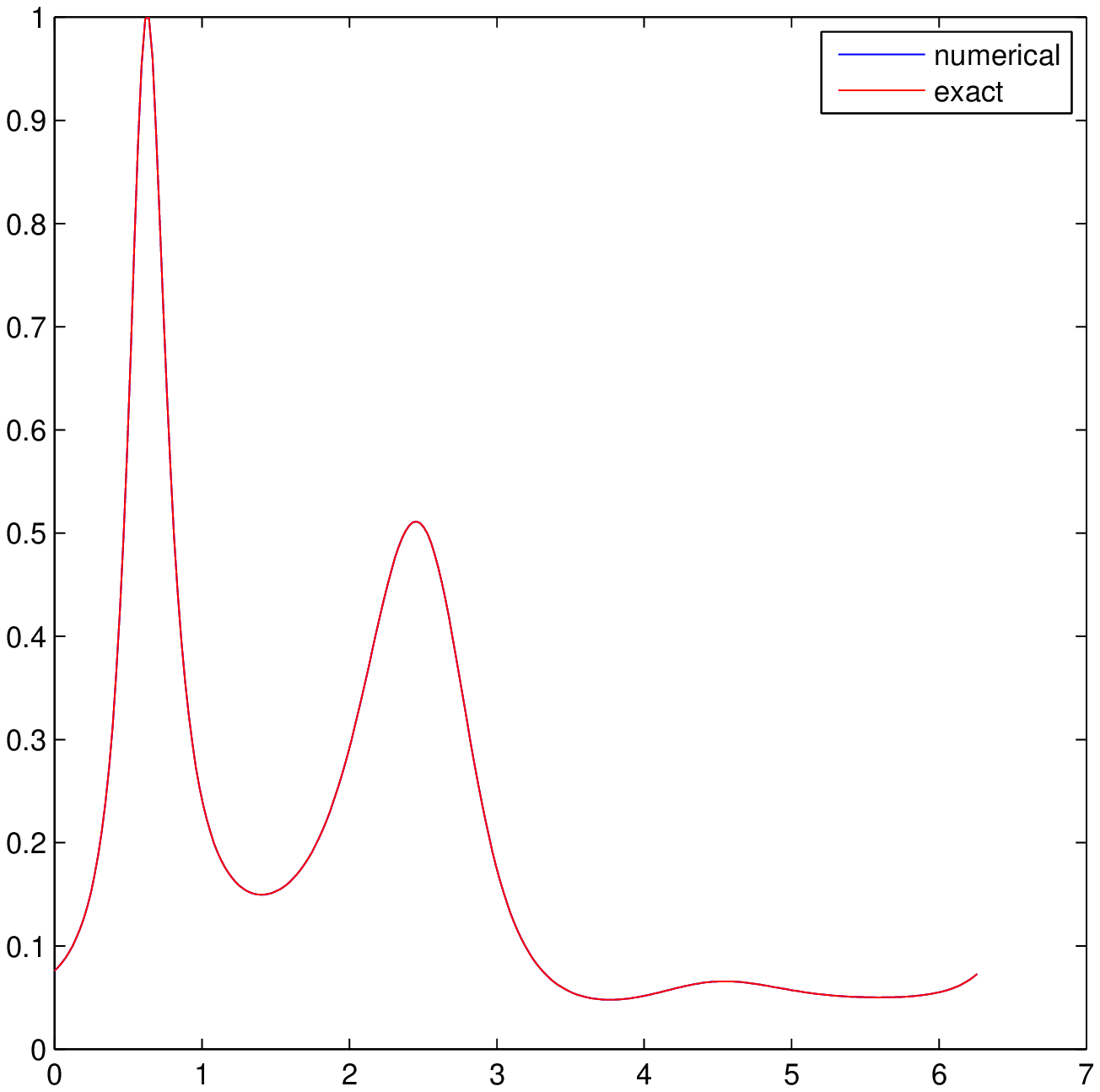}
     \end{center}
    \caption{Upper: the original data in Example 1; Bottom: The shape function obtained by our method (blue) and the exact shape function (red).\label{fig:ex1}}
\end{figure}

\begin{figure}
    \begin{center}
      \includegraphics[width=0.3\textwidth]{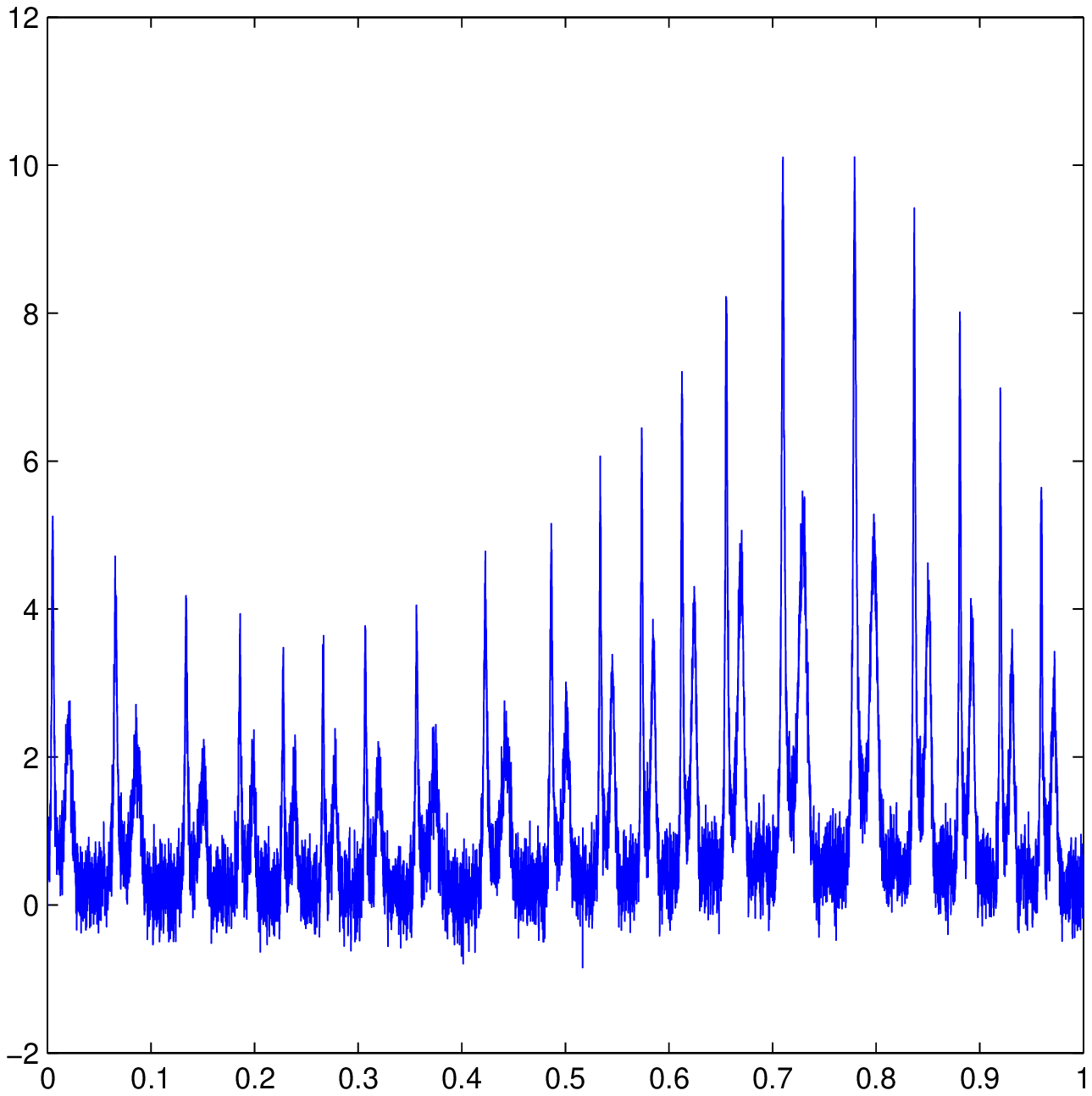}
      \includegraphics[width=0.3\textwidth]{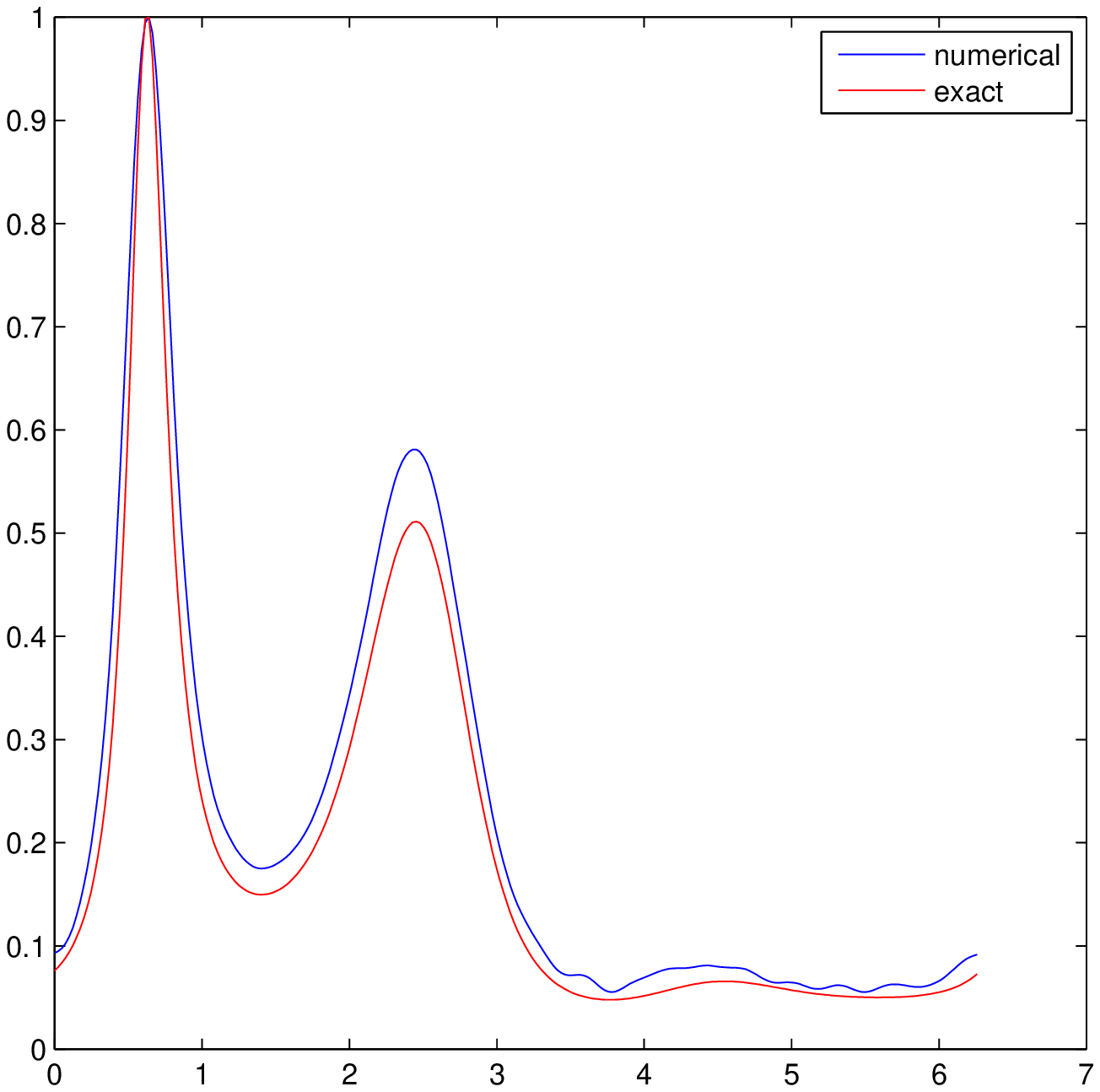}
     \end{center}
    \caption{Upper: The noised data $f(t)+0.3X(t)$, where $f(t)$ is given in Example 1 
and $X(t)$ is the white noise with standard derivative $\sigma^2=1$; Bottom: The shape function obtained by our method (blue) and the exact shape function (red). \label{fig:ex1-noise}}
\end{figure}

\vspace{3mm}
\noindent
\textbf{Example 1.} 
The first example is a simple synthetic data which generated by the following formula:
\begin{eqnarray}
  \theta(t)&=&40\pi t +2\cos(6\pi t),\nonumber\\
 a(t)&=&\frac{1}{2+\sin(2\pi t)},\nonumber\\
\label{eqn:ex1}
f(t)&=&\frac{a(t)}{1.1+\cos\left[\theta(t)+\cos(2\theta(t))\right]}.
\end{eqnarray}
From \eqref{eqn:ex1}, the exact frequency is $\theta'(t)-2\theta'(t)\sin(2\theta(t))$ which has two times more oscillations than 
the original signal $f(t)$. If we relax the shape function to any periodic function 
not necessarily cosine, then the phase function $\theta$ is much smoother than $f(t)$ and the shape function has the form $\cos (t+\cos(2t))$.
In Fig. \ref{fig:ex1}, the shape function given by our method is shown. We can see, in this case, our method could capture the shape function 
very accurately. 
Next, we add Gaussian noise, $0.3X(t)$, to the clean signal to test the robustness of our method. $X(t)$ is the standard Gaussian noise. The 
result is shown in Fig. \ref{fig:ex1-noise}. Even with noise, our method still recovers the shape function with reasonable accuracy.

\vspace{3mm}
\noindent
\textbf{Example 2.}
The second example is the solution of the Duffing equation.
This is an example to demonstrate the importance of the 
intra-wave frequency modulation in some complex dynamic system. 

\begin{figure}
    \begin{center}
\includegraphics[width=0.3\textwidth]{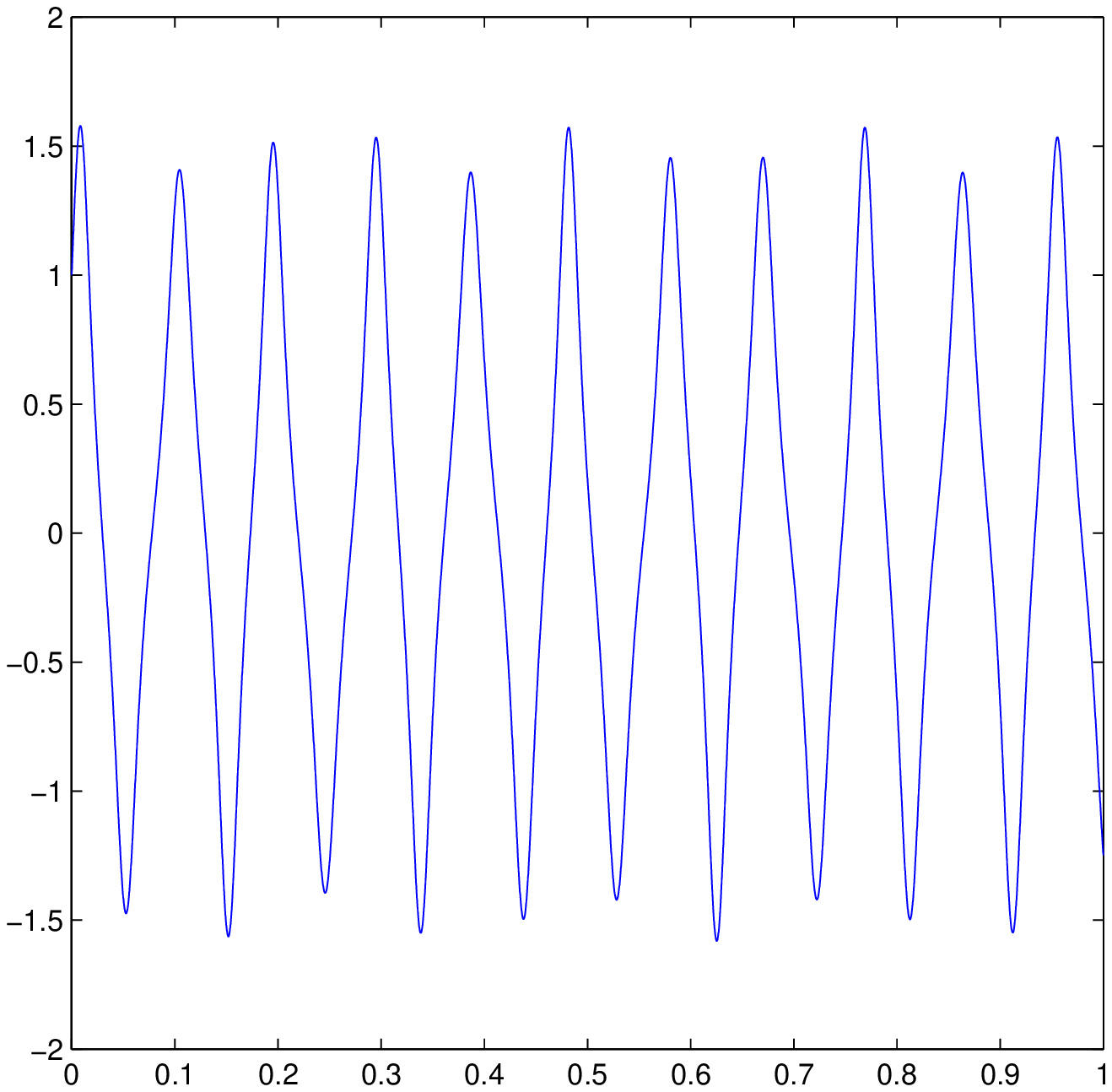}
\includegraphics[width=0.3\textwidth]{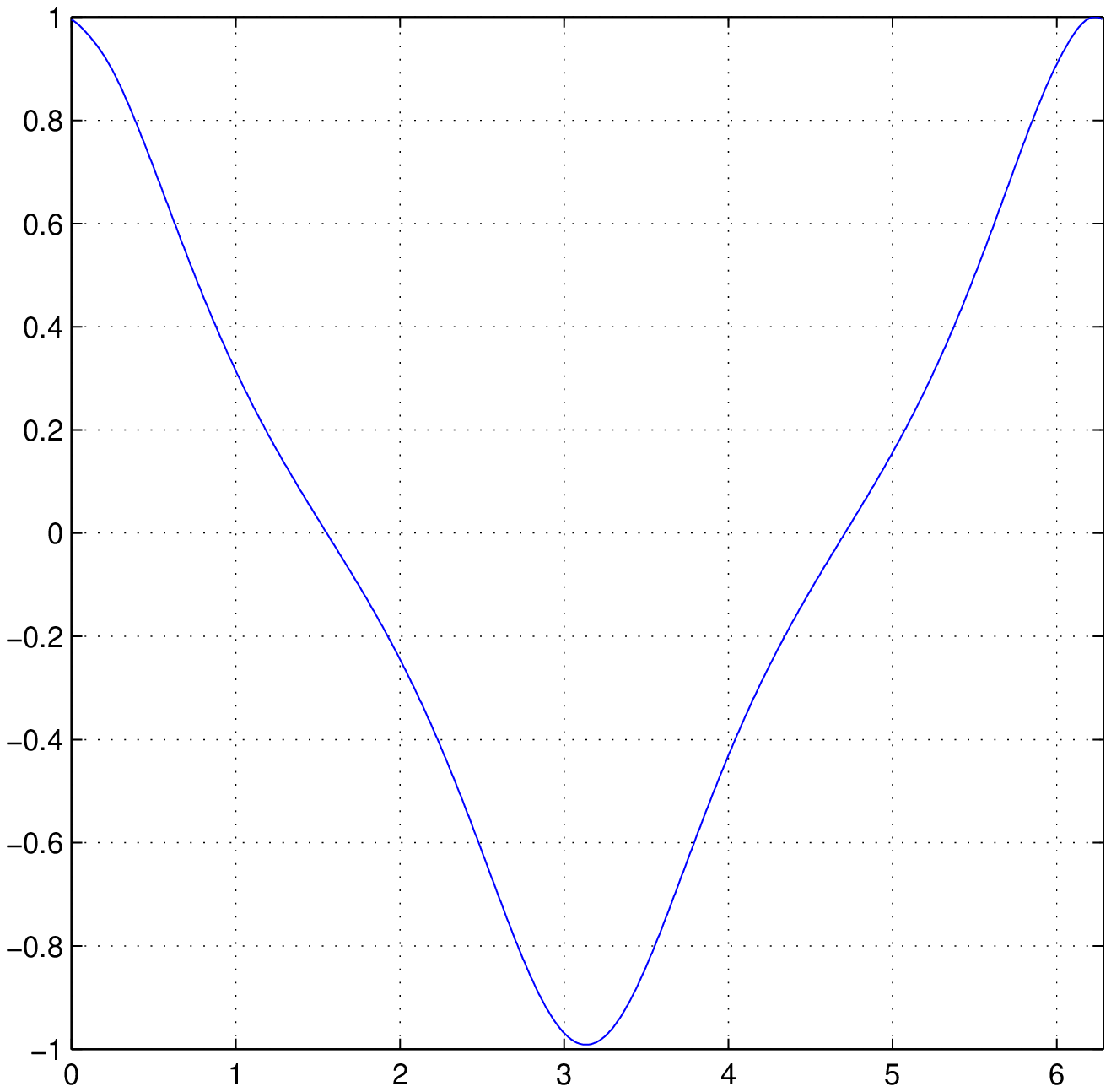}
\includegraphics[width=0.3\textwidth]{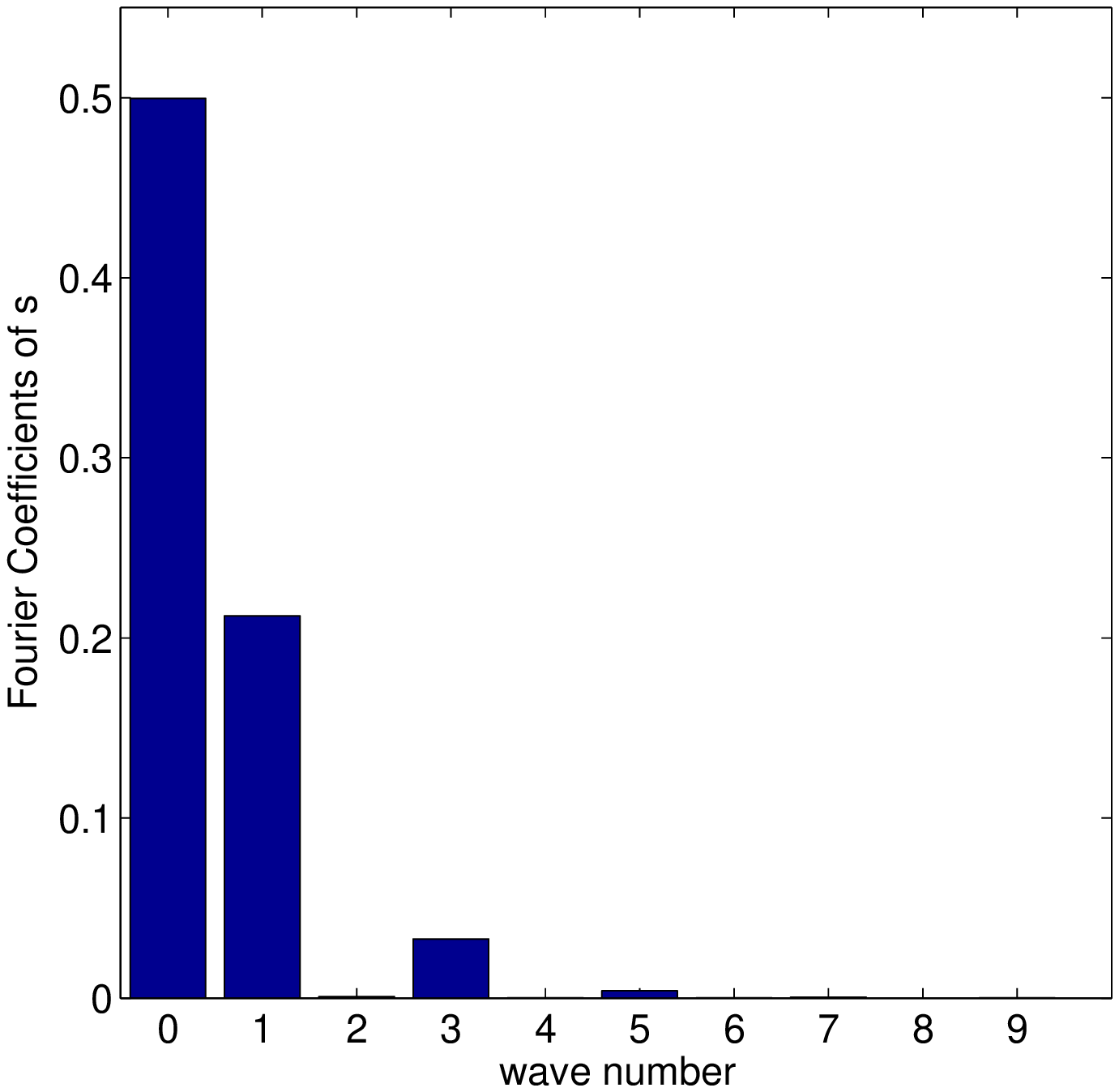}
     \end{center}
    \caption{ \label{duffing}Upper: the solution of duffing equation;
Middle: the shape function $s$; Bottom: the Fourier coefficients of $s$.}
\end{figure}

The Duffing equation is a nonlinear ODE which has the following form:
\begin{eqnarray}
  \label{duffing-1}
  \frac{d^2u}{dt^2}+u+\e u^{1+\omega}=\gamma\cos(\beta t).
\end{eqnarray}
The parameters, $\e,\gamma,\omega$, that we use here to generate the 
solution in Fig. \ref{duffing}, are the same as those in
the paper \cite{Huang98}, $\e=-1,\,\gamma=0.1,\,\beta=\frac{1}{25}$
and $\omega = 2$. The initial condition is $u(0)=u'(0)=1$.

In Fig. \ref{duffing}, we plot the shape function that we obtain from 
the solution of the Duffing equation. In \cite{Huang98}, the example of the
Duffing equation was used to demonstrate that the EMD method is capable of capturing the 
intra-wave frequency modulation. In that computation, the shape function is actually 
fixed to be the cosine function and the intra-wave oscillation is reflected in the instantaneous frequency. 
In our method, since the intra-wave oscillation is absorbed in the shape function, the instantaneous 
frequency is very smooth, but the shape function is not the simple cosine function any more, see
Fig. \ref{duffing}. In this example, we can also express
$s(\theta)$ in terms of $\cos\widetilde{\theta_k}$, from which
we can recover the instantaneous frequency with intra-wave modulation. More 
interestingly, from the Fourier coefficients of the shape function, we can see
the Fourier coefficient is 0 when the wavenumber $k=2$. This actually implies that 
$\omega=2$ in the Duffing equation if we know that the signal comes from an ODE of the particular form
given in \eqref{duffing-1}. This phenomena may be very special, but it suggests that the some quantities, such as the 
deviation of the Fourier coefficients of the shape function, may reflect some important feature of the shape function.

\begin{figure}
    \begin{center}
\includegraphics[width=0.3\textwidth]{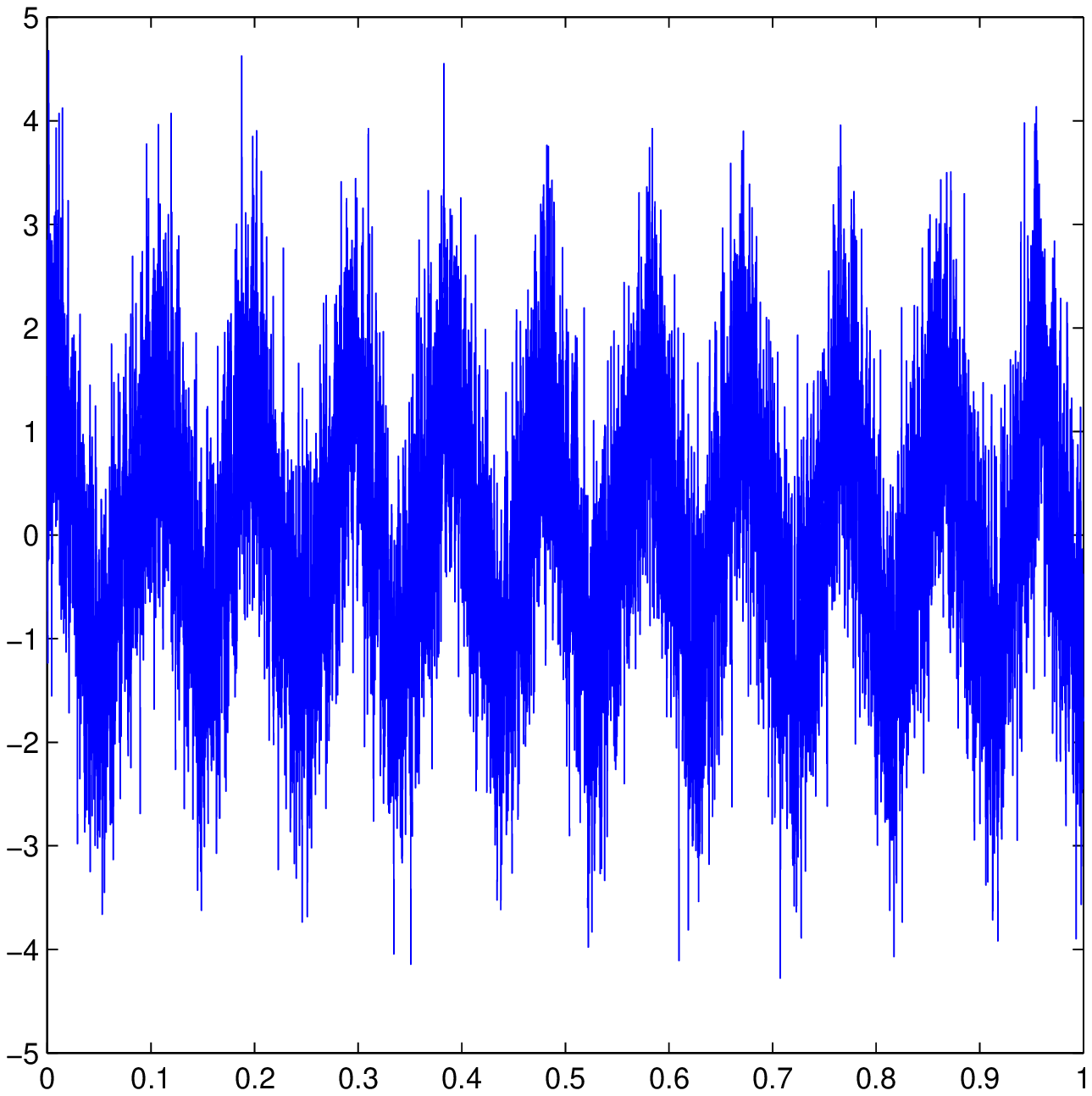}
\includegraphics[width=0.3\textwidth]{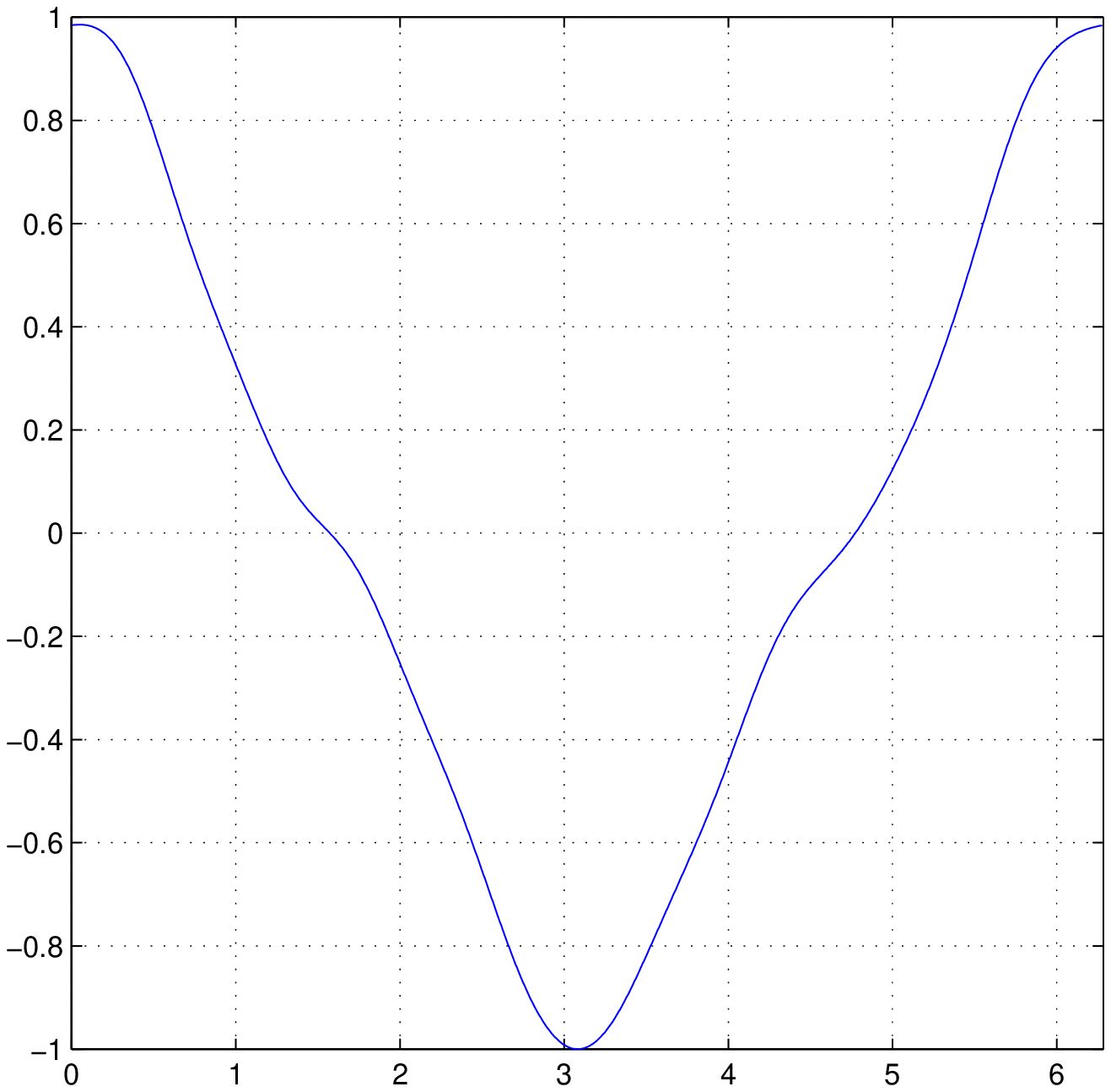}
\includegraphics[width=0.3\textwidth]{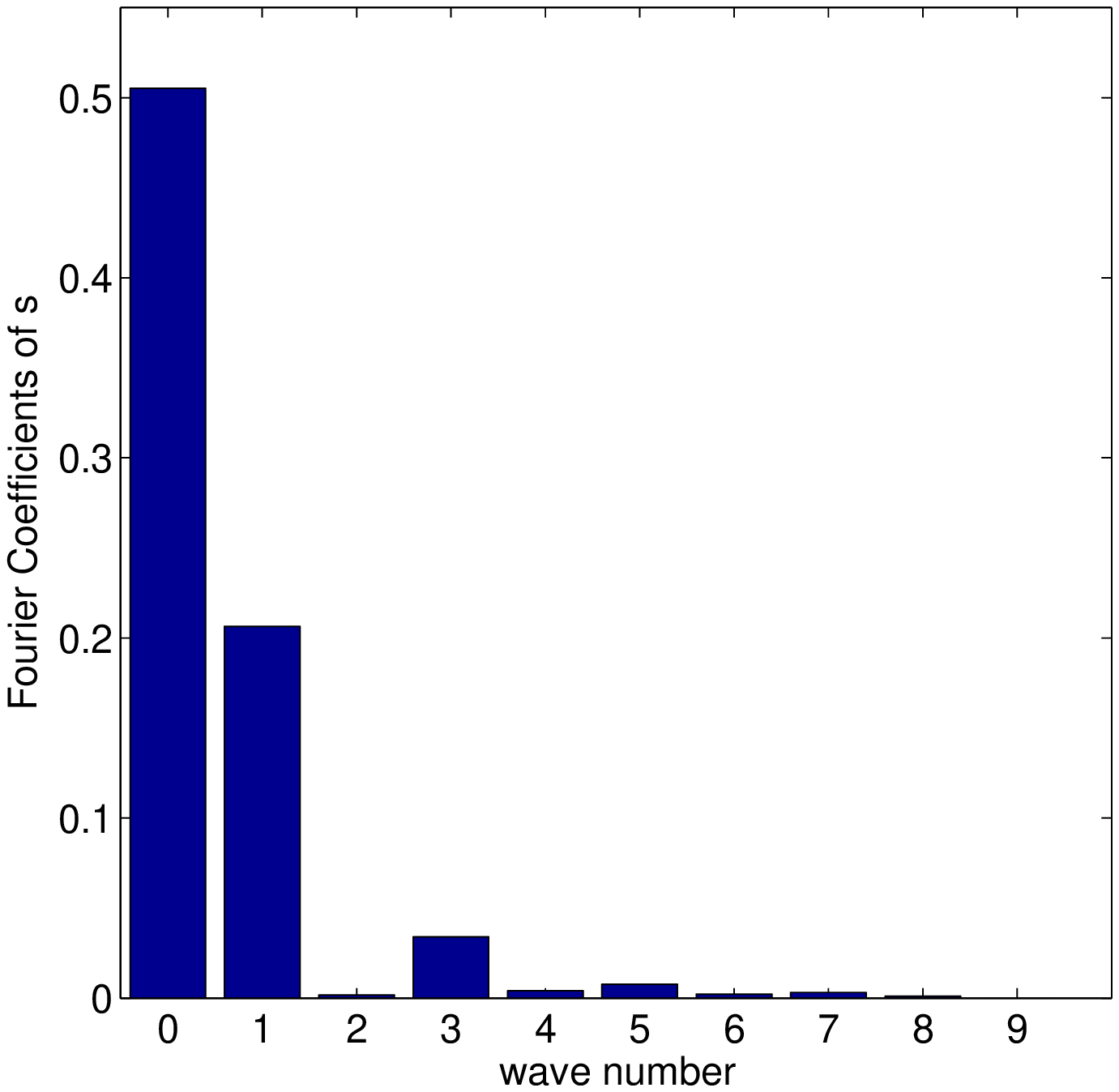}
     \end{center}
    \caption{ \label{duffing-noise}Upper: the solution of duffing equation with noise $X(t)$;
Middle: the shape function $s$; Bottom: the Fourier coefficients of $s$.}
\end{figure}

We also add Gaussian noise $X(t)$ with variance 
$\sigma^2=1$ to the original solution of the Duffing equation.
Fig. \ref{duffing-noise} shows the corresponding results. We can see 
that the shape function extracted from the noisy signal still keeps 
the main characteristics of the shape function extracted from the 
signal without noise. We can also clearly extract the
degree of nonlinearity, $\omega = 2$, even with such large noise
perturbation to the solution of the Duffing equation.

\begin{figure}
    \begin{center}
      \includegraphics[width=0.3\textwidth]{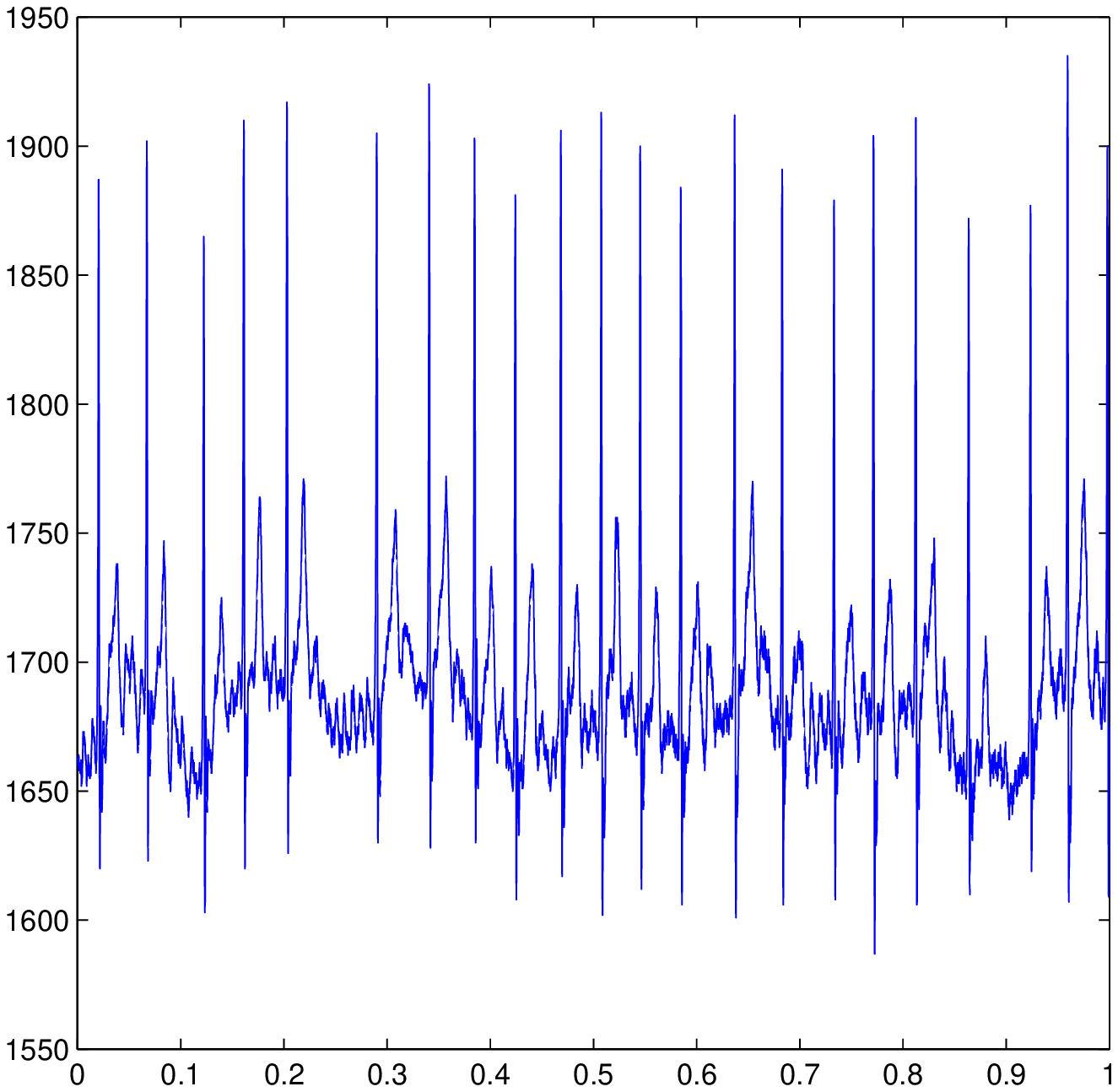}
      \includegraphics[width=0.3\textwidth]{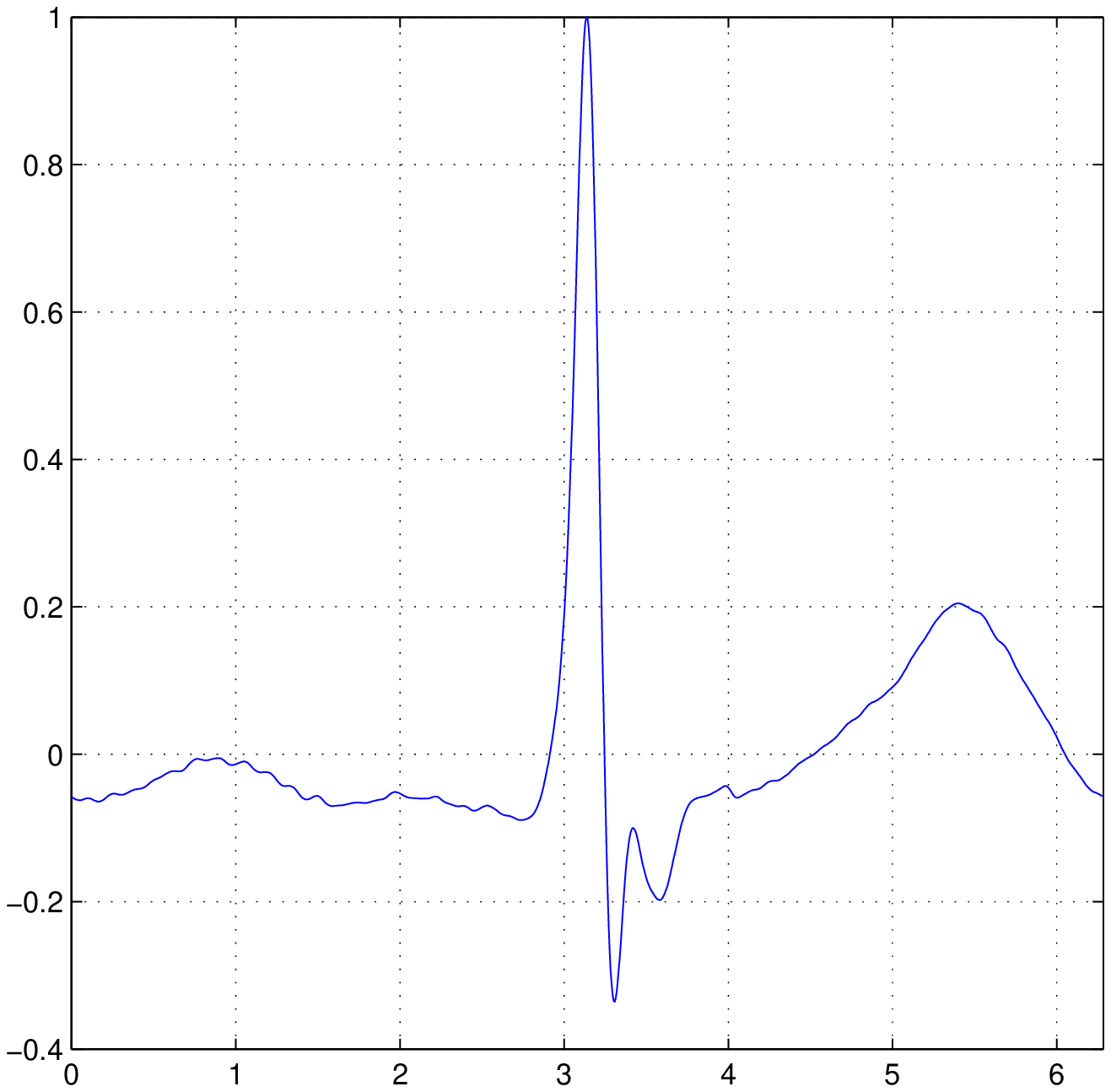}
     \end{center}
    \caption{\label{ECG} Left: The original ECG data; Right: The shape function obtain by our method for the ECG data.}
\end{figure}

\textbf{Example 3: ECG data}
The last example is a piece of electrocardiogram (ECG) data. The length of the data used here is 16s. The bottom picture of Fig. \ref{ECG} 
shows the shape function extracted from this set of ECG signal. We want to remark that it is challenging to extract the shape function from ECG
data since it has sharp peaks in each period. This means that the shape function is not regular and needs many Fourier coefficients 
to well represent it. The shape function that we extracted seems to have all the characteristics of a typical ECG period. 
The interpretation of the significance of the shape function requires expertise in medicine and is beyond our expertise. 

\section{Concluding Remarks}

In this paper, we present an effective and efficient method to extract the shape function from the signal with intra-wave frequency modulation 
by exploiting the intrinsic low rank structure of the data. The current method works only for those signal with one dominated shape function. 
Extracting shape functions for signals with multiple shape functions is much more involved and requires more efforts. How to define an effective index, such 
as the degree of nonlinearity, to reflect the main characteristic of the shape function is another interesting problem, a topic for our future study.

\vspace{0.2in}
\noindent
{\bf Acknowledgments.}
This work was supported by NSF FRG Grant DMS-1159138, DMS-1318377, an AFOSR MURI Grant FA9550-09-1-0613 and a DOE grant DE-FG02-06ER25727. 
The research of Dr. Z. Shi was supported by a NSFC Grant 11201257.

\bibliographystyle{plain}
\bibliography{EMD}

\end{document}